\documentclass[useAMS,usenatbib]{mn2e}

\usepackage{graphicx}
\usepackage{txfonts}

\newcommand{\msun}{M$_{\odot}$}

\title{Modelling self-pollution of globular clusters from AGB stars}
   \author[Y. Fenner, S. Campbell, A. I. Karakas, J. C. Lattanzio and
   B. K. Gibson]{Y. Fenner$^{1}$\thanks{E-mail:
   yfenner@astro.swin.edu.au }, S. Campbell$^{2}$,
   A. I. Karakas$^{2,3}$, J. C. Lattanzio$^{2}$ and
   B. K. Gibson$^{1}$\\ $^{1}$Centre for Astrophysics \&
   Supercomputing, Swinburne University of Technology, Melbourne,
   Australia\\ $^{2}$Centre for Stellar and Planetary Astrophysics,
   Monash University, Melbourne, Australia\\ $^{3}$ICA, Department of
   Astronomy \& Physics, Saint Mary's University, Halifax, Canada }
\begin{document}

\date{}

\pagerange{\pageref{firstpage}--\pageref{lastpage}} \pubyear{2003}

\maketitle

\label{firstpage}

\begin{abstract}    
     A self-consistent model of the chemical evolution of the globular
     cluster NGC~6752 is presented to test a popular theory that
     observed abundance anomalies are due to ``internal pollution''
     from intermediate mass asymptotic giant branch stars. We
     simulated the chemical evolution of the intracluster medium under
     the assumption that the products of Type II SNe are completely
     expelled from the globular cluster, whereas the material ejected
     from stars with $m \lesssim 7$~M$_{\odot}$ is retained, due to
     their weak stellar winds. By tracing the chemical evolution of
     the intracluster gas we have tested an internal pollution
     scenario, in which the Na- and Al-enhanced ejecta from
     intermediate mass stars is either accreted onto the surfaces of
     other stars, or goes toward forming new stars. The observed
     spread in Na and Al was reproduced, but not the O-Na and Mg-Al
     anticorrelations. In particular, neither O nor Mg are
     sufficiently depleted to account for the observations.  We
     predict that the Mg content of Na-rich cluster stars should be
     overwhelmingly dominated by the $^{25,26}$Mg isotopes, whereas
     the latest data shows only a mild $^{26}$Mg enhancement and no
     correlation with $^{25}$Mg.  Furthermore, stars bearing the
     imprint of intermediate mass stellar ejecta are predicted to be
     strongly enhanced in both C and N, in conflict with the empirical
     data. We show that the NGC~6752 data are not matched by a model
     incorporating detailed nucleosynthetic yields from asymptotic
     giant branch stars.  Although these stars do show the hot
     Hydrogen burning that seems required to explain the observations,
     this is accompanied by Helium burning, producing primary C, N, Mg
     and Na (via HBB) which do not match the observations.  Based on
     current theories of intermediate mass stellar nucleosynthesis, we
     conclude that these stars are not responsible for the most of the
     observed globular cluster abundance anomalies.
\end{abstract} 

      

\section{Introduction}

Most Galactic globular clusters (GCs) consist of stars that share the
same characteristic Fe abundance and are consequently considered
monometallic. However, recent studies have found broad variations in
the abundance of some lighter metals. There is a growing list of
globular clusters known to have CNO inhomogeneities and O-Na and Mg-Al
anticorrelations in stars at different stages of evolution, from the
red giant branch to the main-sequence. These abundance anomalies are
not seen in field stars at the same metallicity -- they are peculiar
to the cluster environment.  The two main competing hypotheses to
explain the curious chemical properties of globular cluster stars are:
1) deep mixing, involving processes occurring within stellar interiors
during the course of their evolution; and 2) chemical pollution from
previous stellar generations.

The deep mixing mechanism, which brings freshly processed material to
the surface, is expected to operate in evolved low and intermediate
mass stars (IMS).  Support for this hypothesis comes from the
decreasing C abundance during the red giant branch (RGB) ascent
(Grundahl et~al. 2002).  Temperatures in stars below the turnoff are
not high enough to synthesise Na and Al, nor do they possess a deep
convective layer. Thus the detection of O-Na and Mg-Al
anticorrelations in less evolved stars (Gratton et al.  2001; Grundahl
et al. 2002; Yong et al. 2003) suggests that these chemical patterns
were already present in the gas from which these stars formed, or in
gas that accreted onto the surface of cluster stars. The most popular
candidates for introducing Na- and Al-rich, O- and Mg-poor gas into
the cluster are 4$-$7~{\msun} intermediate mass stars (IMS) on the
asymptotic giant branch (AGB). Cottrell \& Da Costa (1981) described
one variety of AGB pollution whereby GC stars are born from gas
contaminated by AGB ejecta.  Another possibility involves the
accretion of AGB matter into the atmospheres of existing stars (e.g.
D'Antona et al.  1983, Jehin et al. 1998, Parmentier et al.  1999).

We have tested the AGB pollution scenario by calculating the chemical
evolution of globular cluster NGC~6752 using a two stage formation
scenario similar to the model developed by Parmentier et al. (1999).
Parmentier et al. investigated primarily the \emph{dynamical}
evolution of globular clusters, finding that protoclusters can sustain
a series of SNe~II explosions without being disrupted. They assumed
two distinct evolutionary episodes. Firstly, the birth of stars from a
primordial gas cloud leads to a SNe~II phase whose expanding shell
sweeps up the interstellar medium (ISM) into a dense layer. This
triggers the formation of a second generation of stars. In the second
phase, intermediate mass stars from the second stellar generation lose
their envelopes due to stellar winds. This nuclear-processed material
can then be accreted onto the surfaces of lower mass stars.

Thoul et al. (2002) further explored this scenario by calculating the
amount of AGB ejecta retained by globular clusters as well as the
efficiency with which low mass stars can accrete the intracluster
material. They estimate that as much as 95\% of the gas in NGC~6752
has been accreted by cluster stars. They arrive at this value by
assuming that stars have chaotic motions and spend 20\% of their time
in the central homogeneous reservoir of gas. It should be noted that
their results are quite sensitive to the choice of core radius and
stellar velocity, for which they adopted present-day values.
Furthermore, they assume that the only method for gas-removal is via
tidal sweeping as the cluster crosses the Galactic plane. On the other
hand, Smith (1999) proposed that main-sequence stars may generate
winds capable of sweeping all the ejecta from globular clusters,
provided that the line-of-sight velocity dispersion is less than
22~km/s.

Denissenkov \& Herwig (2003) have recently modelled the surface
abundance of a single metal-poor 5~{\msun} star and found that it is
very difficult to simultaneously deplete O and enhance Na.
Furthermore, they found that excess Na is correlated with higher
$^{26}$Mg \emph{and} $^{25}$Mg abundances, whereas Yong et~al. (2003)
only observed a correlation between Na and $^{26}$Mg in NGC~6752.
Given the fine-tuning required to deplete O, and the conflict between
the predicted surface composition of the heavy Mg isotopes and the
data, Denissenkov \& Herwig (2003) question the role of hot-bottom
burning (HBB) during the thermally-pulsing AGB phase in imposing the
anticorrelations. In another recent study, Denissenkov \& Weiss (2004)
point out that the combination of subsolar [C/Fe] and [O/Fe] and very
high [N/Fe] observed in GC subgiants cannot be explained by the
operation of HBB in AGB stars. We reach similar conclusions to these
authors, but based on a fully self-consistent chemical evolution model
of NGC~6752 using yields from 1.25~$-$~6.5~{\msun} stars with initial
compositions determined by the products of population~III stars.

We present a globular cluster chemical evolution (GCCE) model that
examines in detail whether the impact of AGB stellar winds on the
chemical composition of the intracluster medium is consistent with
empirical constraints. This model doesn't discriminate between the
case where clusters stars form from gas already contaminated by AGB
ejecta or the case where existing stars accrete AGB
material\footnote{In the latter case, our model gives the composition
of the accreted matter only. When the star ascends the giant branch,
and experiences the first dredge-up, the (polluted) envelope is mixed
with the (original) interior to produce a different surface
composition.  This is not likely to alter our conclusions,
however.}. The details of the model are described in
Section~\ref{themodel}. The theoretical predictions are shown in
Section~\ref{results} and compared with observations. Finally, the
relevance of this study to the understanding of globular cluster
formation and evolution is discussed in Section~\ref{conclusions}.


\section{The chemical evolution model}\label{themodel}

The chemical evolution of NGC~6752 was predicted using a two stage
formation model.  The first stage traced the chemical evolution of
initially primordial gas following the formation of population III
stars. This stage effectively acted as a prompt initial enrichment,
bringing the gas up to a metallicity of [Fe/H]~=~-1.4 (Gratton et
al. 2001 reported a cluster metallicity of [Fe/H]~=~-1.42, while Yong
et al. 2003 found [Fe/H]~=~-1.62) and imprinting it with the signature
of pop~III ejecta. During the second stage, globular cluster stars
formed from this low-metallicity, $\alpha$-enhanced gas.

\subsection{First stage: Initial enrichment}\label{1ststage}

The conversion of primordial gas into stars was governed by a bimodal
initial mass function (IMF) (Nakamura \& Umemura 2001) favouring the
formation of massive stars. Star formation occurred in a single burst,
with newly synthesised elements being returned on timescales
prescribed by mass dependent lifetimes (Gusten \& Mezger 1982). The
set of zero-metallicity yields from Chieffi \& Limongi (2002) were
used to calculate the enrichment from stars with mass
13~$<$~$m$/M$_{\odot}$~$<$~80. For more massive stars, yields were
given by Umeda \& Nomoto (2002), which cover the mass range 150~$<
m/$M$_{\odot} <$~270. Yields for intermediate and low mass stars were
from the code described in Karakas \& Lattanzio (2003), supplemented
with yields for additional isotopes as well as unpublished yields for
metallicity Z=0.0001. It should be emphasised that intermediate mass
stars played a negligible role in shaping the abundance pattern
imposed by the initial enrichment stage, due to the ``top-heavy'' IMF.

\subsection{Second stage: Globular cluster formation and evolution}\label{2ndstage}

A second generation of stars was then formed from the [Fe/H]~=~-1.4
gas enriched by the Population~III burst. It has been suggested that
expanding shells from the first Type~II SNe may trigger further star
formation (e.g. Thoul et al. 2002). This model adopted a timescale of
10$^7$~yr for the second epoch of star formation. A standard Kroupa,
Tout, \& Gilmore (1993) IMF was adopted, however we assumed that the
globular cluster retains the ejecta from stars with $m$
$\le$~6.5~M$_{\odot}$. Thus, only the yields from intermediate mass
stars impact upon the chemical evolution of the intracluster medium
during this second stage. No contribution from Type~Ia SNe was
included in this GCCE model. The observed uniform iron abundance
implies that SNe~Ia cannot have polluted the cluster during the second
formation stage, while their relatively long characteristic timescales
precludes them from participating during the first phase of
enrichment. We imposed a high efficiency for converting gas into
stars, such that at the end of our simulation the chemical composition
is a blend of roughly 1 part Pop~III enriched gas to 3 parts AGB
ejecta.  Different assumptions will give different mixtures, but will
not alter our main conclusions.

\subsection{Chemical yields from AGB stars}\label{agbyields}

A grid of yields in the mass range 1.25~-~6.5~M$_{\odot}$ was
specifically calculated for this investigation by Campbell et~al.
(2004) using the Mount Stromlo Stellar Structure code (described in
Frost \& Lattanzio 1996; Karakas \& Lattanzio 2003).
Table~\ref{yield_table} presents the total mass (in solar mass units)
of C, N, O, Na, Al and the Mg isotopes released from a star during its
lifetime, as a function of initial stellar mass.

The initial elemental abundances of these detailed stellar models were
imposed by the chemical composition of the gas enriched by the Pop~III
burst. In this way, the chemical evolution model is fully consistent
with the adopted nucleosynthetic prescriptions.

In order to estimate the sensitivity of our results to the
prescriptions used in the stellar evolution calculations, additional
models were run using an alternative: 1) set of reaction rates and 2)
AGB mass-loss prescription -- two of the key factors influencing the
final yields in stellar evolution models. The fiducial set of AGB
yields was calculated using a Reimers (1975) mass-loss law on the RGB
and the Vassiliadis \& Wood (1993) mass-loss law during AGB evolution,
while most of the reaction rates came from the Reaclib Data Tables
(Thielemann, Arnould \& Truran 1991). The reader is referred to
Karakas \& Lattanzio (2003) for greater details of the stellar
models. For comparison with the fiducial set of yields, stellar models
were run for two representative masses, 2.5 and 5.0~M$_{\odot}$, with
the following modifications: 1) ``standard'' reaction
rates\footnote{While most reaction rates in the fiducial AGB models
come from the Reaclib library (Thielemann, Arnould \& Truran 1991),
many of the important rates for producing Mg and Al isotopes have been
updated.  These include: $^{24}$Mg(p,$\gamma$)$^{25}$Al (Powell et
al. 1999); $^{25}$Mg(p,$\gamma$)$^{26}$Al (Iliadis et al. 1996);
$^{26}$Mg(p,$\gamma$)$^{27}$Al (Iliadis et al. 1990);
$^{22}$Ne($\alpha$, n)$^{25}$Mg and $^{22}$Ne($\alpha$,
$\gamma$)$^{26}$Mg (Kaeppeler et al. 1994). } were replaced with NACRE
(Angulo et al. 1999) values for the Ne-Na, Mg-Al and
$^{22}$Ne+$\alpha$-capture chains; and 2) the Vassiliadis \& Wood
(1993) AGB mass-loss law was replaced with the Reimers (1975) law
($\eta_{AGB} = 3.5$). For almost all species considered in this paper,
the final yields predicted using the two sets of reaction rates agree
to within 0.1 dex. The yields are more sensitive to the change in
mass-loss law, as is discussed in Section~\ref{results}. The
efficiency of dredge-up in AGB stars is another crucial factor
influencing the chemical yields. While sensitivity to dredge-up was
not explicitly investigated in the present study, Karakas \& Lattanzio
(2003) tested the effects of reducing the third-dredge-up parameter to
about a third its standard value for the final thermal pulses in the
most massive AGB stars. They found that the yields of $^{23}$Na, C, N
and O changed by only a few percent and $^{25}$Mg, $^{26}$Mg and
$^{27}$Al yields changed by less than 0.15 dex (see their Table 4).  A
more detailed analysis of the dependence on dredge-up will be
published elsewhere.

\begin{table}
\caption{Intermediate Mass Stellar Yields (in solar masses)$^1$ \label{yield_table}}
\centering

\begin{tabular}{l|ccccc}
\hline
 & \multicolumn{5}{c}{Initial stellar mass {\msun} }\\
          &        1.25   &        2.5    &        3.5    &        5.0    & 6.5           \\
\hline
 & \multicolumn{5}{c}{ {\tiny  }  }\\
        C & 8.0x10$^{-4}$ & 1.9x10$^{-2}$ & 1.6x10$^{-2}$ & 9.1x10$^{-3}$ & 7.5x10$^{-3}$ \\
        N & 4.7x10$^{-5}$ & 2.6x10$^{-3}$ & 3.6x10$^{-4}$ & 6.7x10$^{-2}$ & 4.0x10$^{-2}$ \\
        O & 6.8x10$^{-4}$ & 2.1x10$^{-3}$ & 3.0x10$^{-3}$ & 2.1x10$^{-3}$ & 1.4x10$^{-3}$ \\
$^{24}$Mg & 3.4x10$^{-5}$ & 1.0x10$^{-4}$ & 1.5x10$^{-4}$ & 8.6x10$^{-5}$ & 1.2x10$^{-5}$ \\
$^{25}$Mg & 5.0x10$^{-8}$ & 1.6x10$^{-5}$ & 3.5x10$^{-5}$ & 5.5x10$^{-4}$ & 4.4x10$^{-4}$ \\
$^{26}$Mg & 4.0x10$^{-8}$ & 1.2x10$^{-5}$ & 6.1x10$^{-5}$ & 1.4x10$^{-3}$ & 6.1x10$^{-4}$ \\
       Na & 4.3x10$^{-7}$ & 2.0x10$^{-5}$ & 2.4x10$^{-6}$ & 8.7x10$^{-4}$ & 7.8x10$^{-5}$ \\
       Al & 3.0x10$^{-7}$ & 1.1x10$^{-6}$ & 3.1x10$^{-6}$ & 6.3x10$^{-5}$ & 5.8x10$^{-5}$ \\
\hline
\hline
\end{tabular}
\begin{list}{}
\item {\scriptsize $^1$ additional AGB models to appear in Campbell et~al. 2004}
\end{list}

\end{table}

\section{Results}\label{results}

\begin{figure}
\centering
\includegraphics[width=8cm]{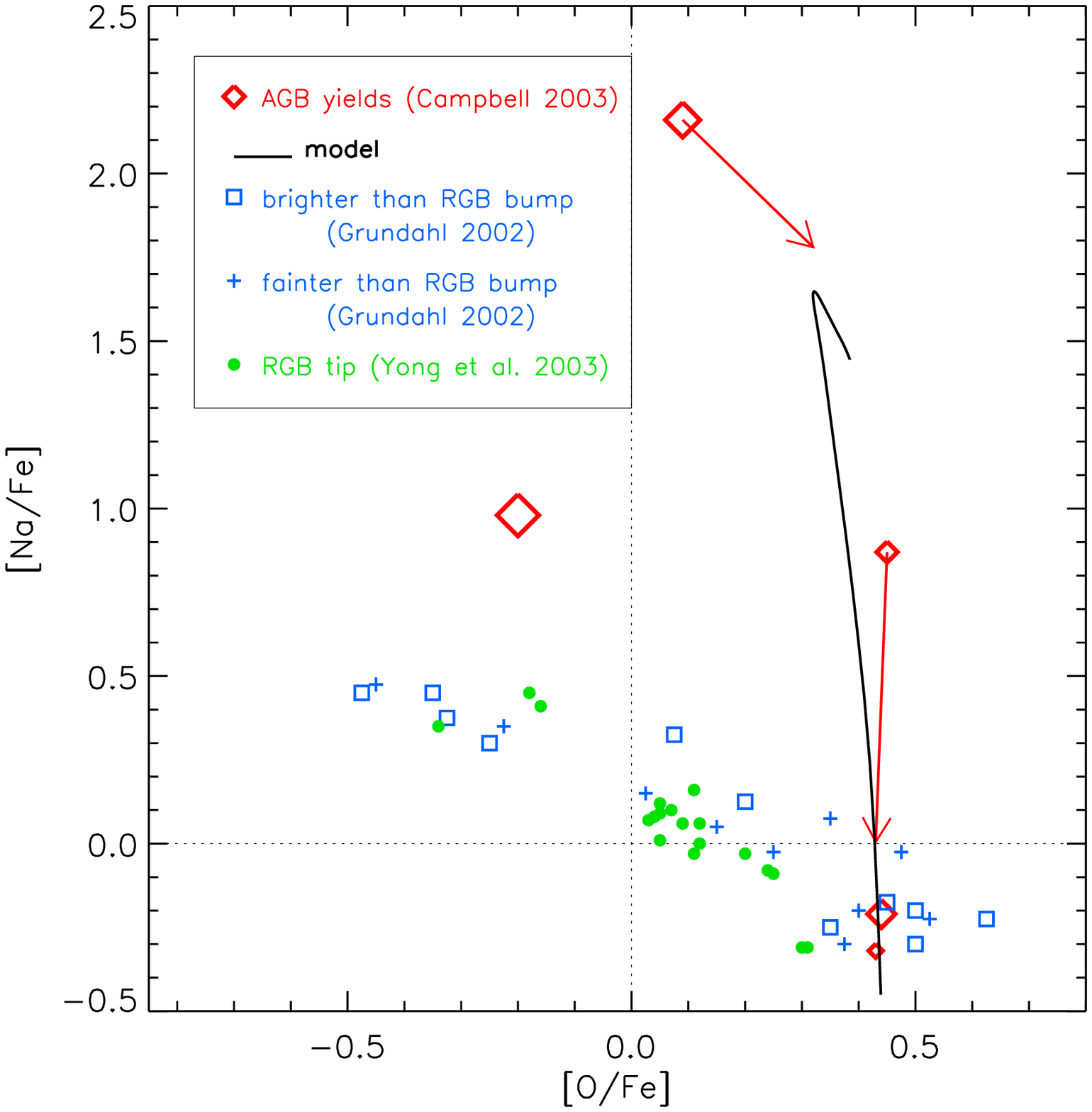}
\includegraphics[width=8cm]{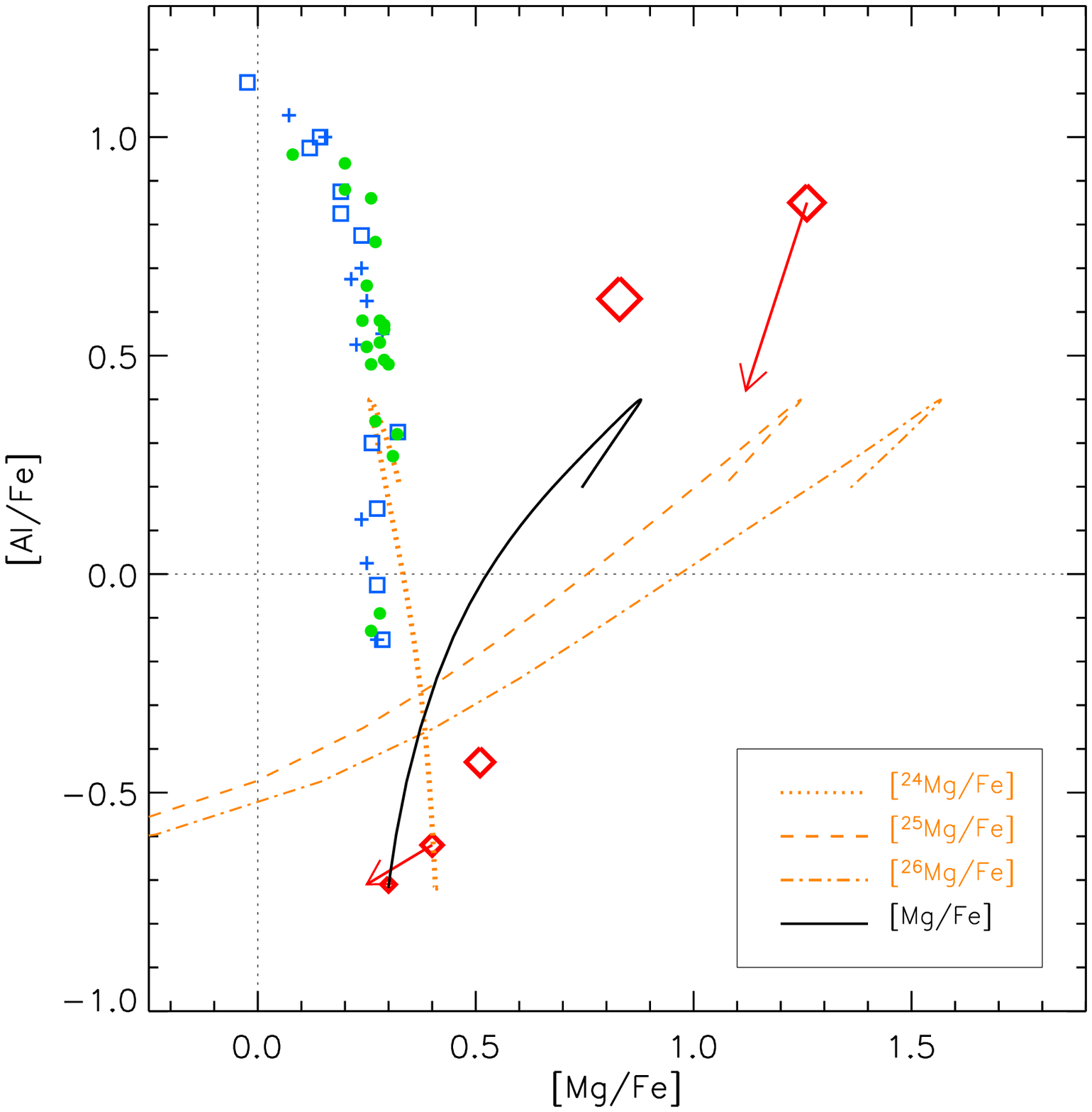}
      \caption{Predicted trend of [Na/Fe] versus [O/Fe] (\emph{a)})
      and [Al/Fe] versus [Mg/Fe] (\emph{b)}) (\emph{thick curve})
      shown against observational data from Grundahl et al. 2002
      (\emph{squares and pluses}) and Yong et al. 2003
      (\emph{circles}). Data from Yong et al. were shifted onto the
      Grundahl et al.  scale.  Diamonds correspond to the 1.25, 2.5,
      3.5, 5.0, and 6.5 {\msun} stellar models of Campbell
      et~al. 2004, where the size of the symbol indicates the stellar
      mass. Arrows indicate the effects of changing the mass-loss law
      for 2.5 and 5.0 {\msun} stars (see text for details). In the
      lower panel, the evolution of [$^{24}$Mg/Fe], [$^{25}$Mg/Fe],
      and [$^{26}$Mg/Fe] are shown by dotted, dashed, and dot-dashed
      lines, respectively. }
      \label{anticorrelations}
\end{figure}
   
Figure~\ref{anticorrelations} compares the observed Na-O (\emph{upper
panel}) and Al-Mg (\emph{lower panel}) anticorrelations in NGC~6752
with model predictions.  Data from Grundahl et al. (2002) and Yong et
al. (2003) show Na-O and Al-Mg anticorrelations in stars that are both
brighter and fainter than the RGB bump. We note that Gratton et
al. (2001) observed similar general abundance trends, but owing to
larger scatter we have omitted their data from this figure. There are
$\sim$~1~dex spreads measured in O, Na, and Al relative to Fe and a
0.3~dex spread in Mg. The observed anticorrelation between [Na/Fe] and
[O/Fe] is poorly matched by the theoretically predicted chemical
evolution of the NGC~6752 intracluster medium. The enrichment from
Pop~III stars brings the gas initially to the lower right-hand part of
the curve (i.e.  [Na/Fe]~$\sim$~$-$0.45 and [O/Fe]~$\sim$~+0.45). The
composition of the intracluster gas then progresses along the curve
toward increasing Na, as the material processed within IMS is released
through stellar winds. The shape of the curve can be understood by
examining the predicted yields from AGB stars, shown in both panels of
Figure~\ref{anticorrelations} by five diamonds whose size indicates
the corresponding stellar mass (i.e. 1.25, 2.5, 3.5, 5.0, or 6.5
{\msun}).  The 6.5~{\msun} model (\emph{largest diamond}) is the first
star to evolve off the main-sequence and release its processed
material into the cluster environment. Consequently, the theoretical
curve (\emph{solid line}) initially begins to move from the O-rich,
Na-poor region of the diagram toward the 6.5~{\msun} diamond. Over
time, the products of progressively lower mass stars influence the
shape of the curve, bearing in mind that low mass stars greatly
outnumber higher mass stars due to the power law IMF.

Only the most massive (6.5~{\msun}) AGB model is capable of
significantly depleting O on a scale approaching the 1~dex spread in
observations.  However, given that the reduction in [O/Fe] for the
6.5~{\msun} star is still only $\sim$~$-$0.6~dex, even a contrived
situation whereby only 6~-~7~{\msun} stars pollute the intracluster
medium leaves the very subsolar [O/Fe] stars unexplained. Moreover,
not enough mass is ejected by 6~-~7~{\msun} stars to form the number
of low-O stars observed.

The AGB models have no difficulty generating Na. Matter ejected by the
5.0~{\msun} star has [Na/Fe]~$\sim$~2.2 -- almost 500 times higher
than the initial Na abundance -- yet the corresponding O depletion is
a mere $\sim$~0.3~dex. The operation of hot-bottom burning in more
massive IMS is a major production site for Na (as well as Al and Mg
isotopes), and Na is ultimately overproduced by this GCCE model. Since
stars with $m > 4$~M$_{\odot}$ are the chief Na production site, the
predicted Na/Fe could be reduced by imposing a more severe upper mass
limit for retaining stellar ejecta.  6.5~M$_{\odot}$ was taken as the
upper mass limit in this model, however, this value depends on many
uncertain properties such as the concentration of mass within the
globular cluster and stellar wind strength and mass loss. The
predicted Na/Fe could also be reduced if there were increased dilution
of the Na-rich AGB ejecta by the Na-poor Pop~III material.

The problem here is that the sodium produced is \emph{primary}.
Helium burning has produced C which has been dredged into the
envelope. The H shell (and HBB) process this into primary N which then
captures two alphas during the thermal pulse to produce $^{22}$Ne.
Some of this Ne is dredged to the surface where the H shell (and HBB)
turn it into the excess Na seen in Figure~1. The observations demand
some Na, but not the huge amounts seen in the models, and this is due
to the origin of the Na being the C produced by helium burning.

Arrows in both panels of Figure~\ref{anticorrelations} indicate the
effects of changing the mass-loss formalism for the 2.5 and 5.0
{\msun} stellar models. Replacing the ``standard'' Vassiliadis \& Wood
(1993) mass-loss law with Reimers (1975) prescription, as described in
Section~\ref{agbyields}, leads to decreases in [Na/Fe] of roughly
$-0.9$ and $-0.4$ dex for the 2.5 and 5.0 {\msun} models,
respectively. Oxygen is almost unchanged for 2.5 {\msun} while at 5.0
{\msun} [O/Fe] is about 0.2 dex higher in the Reimers mass-loss
case. Only the shift due to mass-loss is plotted because it dominates
over the effect from changing the reaction rates. Sodium yields are
significantly higher with the Vassiliadis \& Wood (1993) mass-loss
law, owing to the increased number of third dredge-up episodes that
progressively increase the Na abundance at the surface, and to the
fact that much of the convective envelope is lost during the final few
thermal pulses, when the surface abundance of Na is at its
highest. Because the rate of mass-loss proceeds more steadily under
the Reimers (1975) law, more material is lost earlier on in the AGB
phase, prior to the high envelope abundance of Na. While the adoption
of Reimers (1975) mass-loss helps remedy the problem of Na
overproduction, it only worsens the oxygen discrepancy. This is
because depletion of $^{16}$O in the stellar envelope is due to HBB,
which operates for a shorter time in models with Reimers mass-loss
(due to the faster initial mass-loss rate as compared to Vassiliadis
\& Wood (1993)). In addition, $^{16}$O is only significantly depleted
in models with $m \gtrsim$ 5 {\msun}, since these models exhibit (high
temperature) HBB.

We stress that a disagreement between the GCCE model and the data
would exist regardless of the precise shape of the IMF. Indeed,
inspection of the individual yields of AGB stars of various masses
reveals that no choice of IMF would reproduce the observations.

The lower panel of Figure~\ref{anticorrelations} shows that this GCCE
model predicts about a 1~dex spread in [Al/Fe]. The predicted spread
is consistent with the star-to-star variation, but the absolute values
are $\sim$~0.6~dex lower than observed. In stark contrast to the
measured Al-Mg anticorrelation, we predict that the total Mg abundance
\emph{increases} with increasing Al. This discrepancy would not be
resolved by any choice of IMF, since none of the yields of individual
stars are depleted in total Mg. Once again, the shape of the
theoretical curve reflects the yields from different mass AGB stars
(\emph{diamonds}) with different lifetimes and the arrows reflect the
effect of changing the mass-loss law. It is evident that both
mass-loss cases lead to a discrepancy with the data. The Pop~III burst
leaves a high-Mg, low-Al chemical signature on the gas from which the
intermediate and low mass stars begin to form. The intracluster gas is
then enriched in both Mg and Al, which are produced by the AGB stars
and expelled through stellar winds. The increase in [Mg/Fe] is
entirely due to the enhanced abundance of the heaviest magnesium
isotopes, $^{25}$Mg and $^{26}$Mg, which are produced primarily in the
He-burning shell of intermediate-mass AGB stars (Karakas \& Lattanzio
2003). The dramatic increase in the heavier Mg isotopes is revealed by
the dashed and dot-dashed lines showing the behaviour of
[$^{25}$Mg/Fe] and [$^{26}$Mg/Fe], respectively. Isolating
[$^{24}$Mg/Fe] (\emph{dotted line}), we recover an anticorrelation
resembling the data, albeit offset to lower [Al/Fe] values.
Hot-bottom burning in the more massive AGB stars is responsible for
this slight depletion of $^{24}$Mg, which is converted into
$^{25}$Mg. However, this only occurs for $m \gtrsim 5$~{\msun}, where
temperatures in the H-shell exceed 90 million K. It should be noted
that the uncertainty in the $^{26}$Mg(p,$\gamma$)$^{27}$Al reaction
rate permits significantly more $^{27}$Al being produced at the
expense of $^{26}$Mg (Arnould, Goriely, \& Jorissen 1999) than in the
present models.

Once again, the problem is the products of helium burning. The primary
$^{22}$Ne mentioned earlier also suffers alpha captures to produce the
heavy Mg isotopes via $^{22}$Ne($\alpha$,n)$^{25}$Mg and
$^{22}$Ne($\alpha$,$\gamma$)$^{26}$Mg Thus the Mg seen in these AGB
stars is again primary, and due to Helium burning. The
anti-correlation seen in many globular clusters between Mg and Al is
totally swamped by the production of Mg in the helium burning region.
Thus our models predict an \emph{increase} in total Mg whereas the
data show a \emph{decrease}.

\begin{figure*}
\centering
\includegraphics[width=17cm]{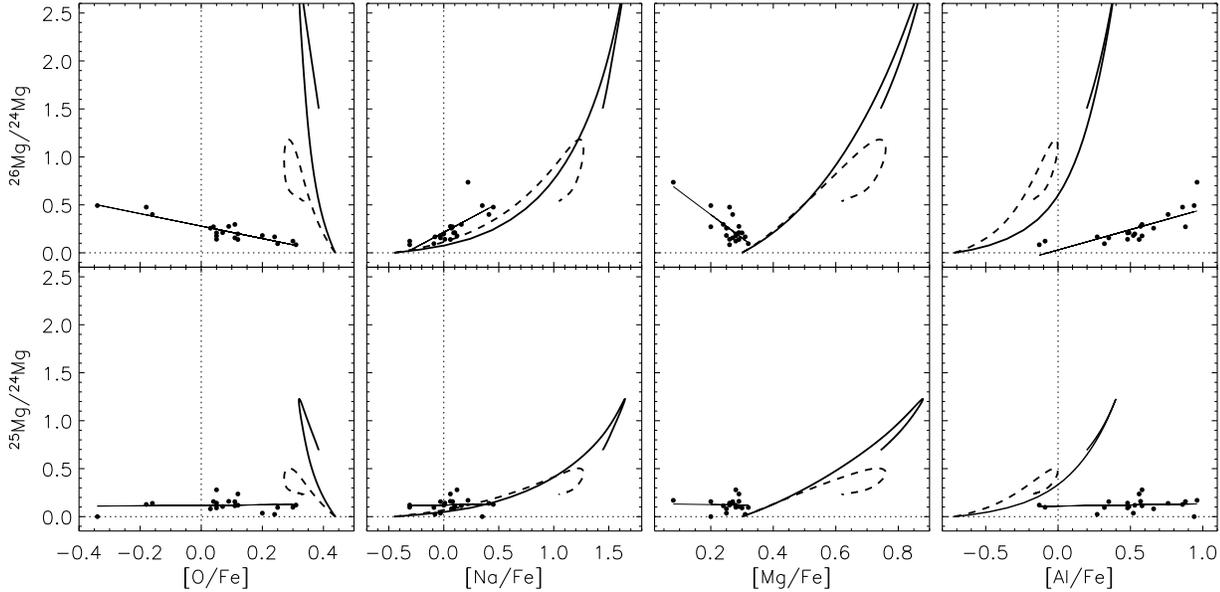}
      \caption{The trend of Mg isotopic ratios with O, Na, Mg, and Al
        abundance predicted by the NGC~6752 model presented in this
        paper. The thick solid lines show predictions for
        $^{25}$Mg/$^{24}$Mg (\emph{top panels}) and
        $^{26}$Mg/$^{24}$Mg (\emph{bottom panels}). Dashed lines show
        predictions from the Reimers (1975) AGB mass-loss model, as
        described in the text. Circles correspond to data from Yong et
        al. 2003 showing positive correlations between
        $^{26}$Mg/$^{24}$Mg and [Na,Al/Fe]; anticorrelations between
        $^{26}$Mg/$^{24}$Mg and [O,Mg/Fe]; and no correlation for
        $^{25}$Mg/$^{24}$Mg. The lines of best fit to the data are
        represented by thin lines.}
         \label{mgisotopes}
\end{figure*}
   
Isotopic ratios have the potential to offer more insight into the
source of anomalous chemical patterns than elemental abundances
alone. Unfortunately, the important reaction rates for the production
of Mg isotopes are beset by uncertainties. In particular, the yield of
$^{26}$Mg/$^{25}$Mg from AGB stars is sensitive to the
$^{22}$Ne($\alpha$,n)$^{25}$Mg and
$^{22}$Ne($\alpha$,$\gamma$)$^{26}$Mg rates, neither of which are
tightly constrained (see Figure 12 from Arnould, Goriely, \& Jorissen
1999).  In Figure~\ref{mgisotopes} we show the variation in Mg
isotopic ratios as a function of O, Na, Mg, and Al abundance. The
circles represent measurements of stars at the tip of the red giant
branch in NGC~6752 by Yong et~al. (2003). The thick solid lines are
predictions from our standard model with Vassiliadis \& Wood (1993)
mass-loss. Predictions from the Reimers (1975) AGB mass-loss model are
indicated by dashed lines (where we interpolated and extrapolated the
uncertainty calculated for the 2.5 and 5 {\msun} stars to other
masses). The behaviour of $^{26}$Mg/$^{24}$Mg and $^{25}$Mg/$^{24}$Mg
is shown in the upper and lower series of panels, respectively. Yong
et~al.  (2003) revealed clear anticorrelations between
$^{26}$Mg/$^{24}$Mg and [O,Mg/Fe] and positive correlations between
$^{26}$Mg/$^{24}$Mg and [Na,Al/Fe].  Conversely, there is no evidence
that $^{25}$Mg/$^{24}$Mg varies with the abundance of these four
elements. Theoretical predictions based on current theories of AGB
nucleosynthesis deviate markedly from the observations in a number of
important ways: 1) Al-rich and O-poor stars are expected to have much
higher ($^{25}$Mg~+~$^{26}$Mg)/$^{24}$Mg ratios than is observed, 2)
the $^{25}$Mg abundance is tightly linked to $^{26}$Mg abundance in
the models, counter to observations, and 3) the predicted trend of
$^{25,26}$Mg/$^{24}$Mg with total Mg abundance (i.e. [Mg/Fe]) moves in
the opposite direction to the data, due to the primary production of
Mg in these AGB stars.  Regarding the first point, we stress that the
overproduction of $^{25,26}$Mg by our GCCE model is not disastrous for
the AGB self-pollution scenario. Lower values of
($^{25}$Mg~+~$^{26}$Mg)/$^{24}$Mg could be obtained by diluting the
AGB ejecta with more of the initial gas. Moreover, the dashed lines
demonstrate that a factor of $\sim$~2 reduction in the AGB yield of
($^{25}$Mg~+~$^{26}$Mg)/$^{24}$Mg can be attained by adopting a
``steadier'' rate of of AGB mass-loss, such as the Reimers (1975)
prescription described in Section~\ref{agbyields}. However, points 2
and 3, listed above, represent more serious problems for the AGB
self-pollution picture, and are robust to the variations in stellar
inputs considered in this study.

\begin{figure}
\centering
\includegraphics[width=8cm]{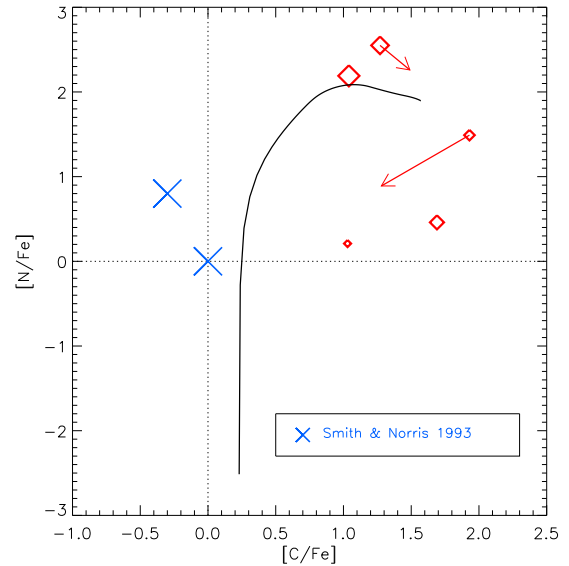}
      \caption{Predicted evolution of [N/Fe] with [C/Fe] (\emph{solid
          line}). Symbols have the same meaning as in
        Figure~\ref{anticorrelations} except that crosses denote data
        from Smith \& Norris 1993. }
         \label{NvsC}
\end{figure}
    
Figure~\ref{NvsC} compares the predicted trend of [N/Fe] versus [C/Fe]
from the AGB pollution model (\emph{solid line}) against observations
of AGB stars in NGC~6752 from Smith \& Norris (1993)
(\emph{crosses}). Diamonds correspond to yields from the AGB models
for five initial masses from 1.25~to~6.5~{\msun}, and arrows reflect
the effect of changing the mass-loss law. Red giants in NGC~6752 have
been found to exhibit bimodal C and N abundances with one group having
roughly solar [C/Fe] and [N/Fe], and the other group being N-rich and
C-poor (e.g. da Costa \& Cottrell 1980; Norris et~al. 1981; Smith \&
Norris 1993). The empirical data points to an anticorrelation between
C and N that has previously been explained in terms of the operation
of a deep mixing mechanism in evolved stars. The dependence of the
molecular CN-band strength on stellar luminosity (e.g. Suntzeff \&
Smith 1991) is readily understood as a result of increased mixing of
N-rich, C-poor material from within the stellar interior to the
surface, as a function of evolutionary stage. More recently, however,
Grundahl et~al. (2002) found an anticorrelation between [C/Fe] and
[N/Fe] in NGC~6752 stars that is independent of luminosity. Cannon
et~al. (1998) found a similar CN bimodality and anticorrelation in 47
Tuc from a sample of stars that included not just giants, but also
unevolved main-sequence stars whose shallow convective layers preclude
dredge-up of CNO-cycled material. The presence of the same CN trend in
both dwarfs and giants led Cannon et al. (1998) to conclude that deep
mixing was not singularly responsible for CN abundance anomalies in 47
Tuc. It seems likely that the CN patterns in NGC~6752 stars arise from
a combination of deep mixing and some form of external pollution.

Figure~\ref{NvsC} casts doubt on AGB stars being a major source of
external pollution.  All the AGB models from 1.25~$-$~6.5~{\msun}
expel material that is enhanced in both N \emph{and} C. From
Figure~\ref{NvsC} it is clear that main-sequence stars severely
polluted by AGB material are expected to also exhibit heightened C and
N abundances. The positive correlation between C and N in the
intracluster gas predicted by our GCCE model fails to match the
empirical trend, however, we note that our calculations only reflect
the chemical evolution of the intracluster medium. The abundance
pattern of the gas at a particular time corresponds to the initial
composition of stars born at that time, whereas the observed abundance
of elements like CNO in evolved red giants are likely to differ from
their starting abundances due to internal synthesis and mixing.

\begin{figure}
  \centering \includegraphics[width=8cm]{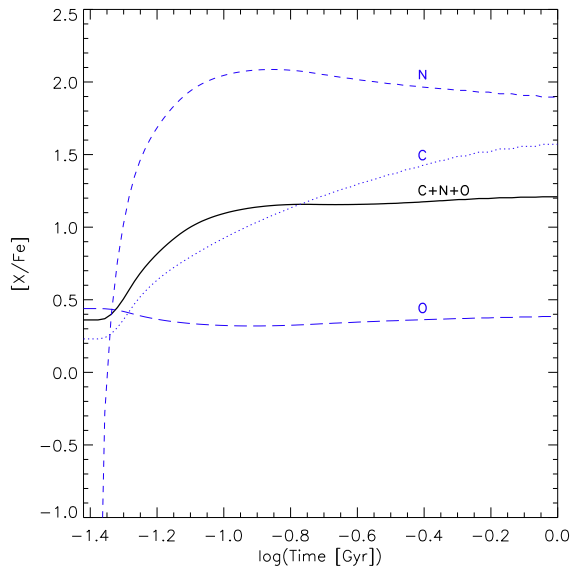}
  \caption{Temporal evolution of [C/Fe] (\emph{dotted line}), [N/Fe]
  (\emph{short dashed line}), [O/Fe] (\emph{long dashed line}), and
  [C+N+O/Fe] (\emph{solid line}). In contrast to observations in
  NGC~6752 and other globular clusters, C+N+O abundance is predicted
  to vary by an order of magnitude if AGB stars are responsible for
  the Na and Al enhancement.}  \label{CNO}
\end{figure}

Figure~\ref{CNO} reveals an almost order of magnitude rise in
[C~+~N~+~O/Fe] in the intracluster gas within 1~Gyr of formation. A
slight drop in the O abundance is more than compensated for by a
dramatic increase in C and N. This robust prediction is based on
intermediate mass stellar nucleosynthesis and poses further
difficulties for the AGB pollution scenario as an explanation for
globular cluster abundance anomalies, since C~+~N~+~O is found to be
approximately constant in many GCs (Ivans et~al. 1999). AGB stars have
been proposed in the literature as promising candidates for producing
the observed GC abundance anomalies because they exhibit the required
hot H burning via hot-bottom burning. However, because these stars
also dredge-up the products of He-burning, C~+~N~+~O is not conserved
in the models, in conflict with the data.

We note that in order to achieve an order of magnitude increase in the
intracluster abundance of Al via the AGB pollution scenario, our
models simultaneously predict an intracluster medium helium mass
fraction, Y, approaching 0.3. This represents a $\sim$~0.05 increase
in Y over the primordial value. D'Antona et~al. (2002) have found that
the effects on stellar evolution due to this level of He enrichment
would be difficult to measure observationally but could lead to
extended blue tails in the horizontal branch morphology.


\section{Conclusions}\label{conclusions}
   
The presence of variations in C, N, O, Mg, Na, and Al in globular
cluster stars yet to ascend the red giant branch provides compelling
evidence for these chemical patterns already being in place in the gas
from which cluster stars formed, or in gas that later polluted their
atmospheres. Otherwise, one would expect to see chemical homogeneity
in stars below a certain luminosity, unless our understanding of deep
mixing is seriously flawed. We have constructed a self-consistent
model of the chemical evolution of the intracluster gas, that included
custom-made detailed stellar models, to test whether the observed
inhomogeneities may be caused by contamination from material processed
by intermediate mass stars during their AGB phase. This model is
compatible with either a scenario in which there is coeval and
stochastic sweeping of the intermediate mass stellar ejecta by
existing lower mass stars, or one in which new stars form from AGB
polluted material. In the latter case, there would be a small age
spread with Na-rich stars being a few hundred million years younger
than Na-poor stars.

We find that, regardless of either the mechanism for polluting cluster
stars with AGB material or the level of dilution of Pop~III material
by AGB ejecta, intermediate mass stars are unlikely to be responsible
for most of the abundance anomalies. While metal-poor AGB models
generate large quantities of Na and Al that may account for the
observed spread in these elements in NGC~6752, the AGB pollution
scenario encounters a number of serious problems: 1) O is not depleted
within AGB stars to the extent required by observations, 2) Mg is
\emph{produced} when it should be \emph{destroyed}, 3) C+ N + O
does not remain constant in AGB processed material, and 4) $^{25}$Mg
is correlated with $^{26}$Mg in the modelled AGB ejecta, conflicting
with the Yong et al. (2003) observations.

Note that all of these problems stem from the addition of helium
burning products into the AGB star ejecta. Perhaps a generation of AGB
stars which experience HBB but almost no dredge-up would fit the data
better!

The model presented in this paper could be generalised for application
to other globular clusters by varying three main parameters: the
initial metallicity; the upper mass limit beyond which stellar winds
are too energetic for the cluster to retain the ejecta (this depends
on the mass concentration and gravitational potential); and the
efficiency and duration of star formation.

\section*{Acknowledgements}

Financial support from the Australian Research Council (ARC) is
gratefully acknowledged. We also acknowledge the Monash Cluster
Computing Laboratory, the Victorian Partnership for Advanced Computing
and the Australian Partnership for Advanced Computing for use of
supercomputing facilities. JCL would like to thank the Institute of
Astronomy and Churchill College, Cambridge University, for their
hospitality and wine cellars, respectively. We thank the referee for
helpful comments on the paper.

\bsp

\label{lastpage}


\begin{thebibliography}{}
\bibitem[1999]{} Angulo C., Arnould M., Rayet M., et al. (NACRE
   collaboration), 1999, Nuclear Physics A 656, 3
\bibitem[Arnould, Goriely, \& Jorissen(1999)]{arnould} Arnould, 
   M., Goriely, S., \& Jorissen, A.\ 1999, A\&A, 347, 572 
\bibitem[2003]{campbell} Campbell, S. W., 2004, in prep
\bibitem[Cannon et al.(1998)]{1998MNRAS.298..601C} Cannon, R.~D., Croke, 
    B.~F.~W., Bell, R.~A., Hesser, J.~E., \& Stathakis, R.~A.\ 1998, MNRAS, 
    298, 601 
\bibitem[2002]{chieffi} Chieffi, A. \& Limongi, M., 2002, ApJ 577, 281
\bibitem[cottrell]{1981} Cottrell, P. L. \& Da Costa, G. S., 1981,
    ApJ, 245L, 79
\bibitem[1983]{dantona} D'Antona, F., Gratton, R. \& Chieffi, A.,
    1983, Mem. Soc. Astron. It., 54, 173
\bibitem[D'Antona et al.(2002)]{2002A&A...395...69D} D'Antona, F., Caloi, 
    V., Montalb{\' a}n, J., Ventura, P., \& Gratton, R.\ 2002, A\&A, 395, 69 
\bibitem[Denissenkov \& Herwig(2003)]{2003ApJ...590L..99D} Denissenkov, 
    P.~A.~\& Herwig, F.\ 2003, ApJL, 590, L99 
\bibitem[Denissenkov \& Weiss(2004)]{2004ApJ...603..119D} Denissenkov, 
    P.~A.~\& Weiss, A.\ 2004, ApJ, 603, 119 
\bibitem[1999]{dinescu} Dinescu, D. L., Girard, T. M. \& van Altena,
    W. F., 1999, AJ 117, 1792
\bibitem[Frost \& Lattanzio(1996)]{1996ApJ...473..383F} Frost, C.~A.~\& 
   Lattanzio, J.~C.\ 1996, ApJ, 473, 383 
\bibitem[2001]{gratton} Gratton, R. G. et al., 2001, A\&A 369, 87
\bibitem[2002]{grundahl} Grundahl, F., Briley, M, Nissen, P. E. \&
    Feltzing, S. 2002, A\&A 385, L14
\bibitem[gusten]{1982} Gusten, R. \& Mezger, P. G., 1982, Vistas
    Astron. 26, 159
\bibitem[Iliadis et al.(1990)]{1990NuPhA.512..509I} Iliadis, C., et al.\ 
   1990, Nucl. Phys. A, 512, 509 
\bibitem[iliadis]{1996}Iliadis, C., Buchmann, L., Endt, P. M., Herndl,
   H., \& Wiescher, M. 1996, Phys. Rev. C, 53, 475
\bibitem[Ivans et al.(1999)]{1999AJ....118.1273I} Ivans, I.~I., Sneden, C., 
   Kraft, R.~P., Suntzeff, N.~B., Smith, V.~V., Langer, G.~E., \& Fulbright, 
   J.~P.\ 1999, AJ, 118, 1273 
\bibitem[1998]{jehin} Jehin, E., Magain, P., Neuforge, C., Noels, A. \&
    Thoul, A. A., 1998, A\&A, 330, L33
\bibitem[Kaeppeler et al.(1994)]{1994ApJ...437..396K} Kaeppeler, F., et 
   al.\ 1994, ApJ, 437, 396
\bibitem[Karakas \& Lattanzio(2003)]{2003PASA...20..279K} Karakas, A.~I.~\& 
  Lattanzio, J.~C.\ 2003, Publications of the Astronomical Society of 
  Australia, 20, 279 
\bibitem[Kraft(1994)]{1994PASP..106..553K} Kraft, R.~P.\ 1994, PASP, 106, 
  553 
\bibitem[Kroupa, Tout, \& Gilmore(1993)]{1993MNRAS.262..545K} Kroupa, P.,
   Tout, C.~A., \& Gilmore, G.\ 1993, MNRAS, 262, 545
\bibitem[2001]{nakamura} Nakamura, F. \& Umemura, M., 2001, ApJ 548,
    19
\bibitem[Norris, Cottrell, Freeman, \& Da Costa(1981)]{1981ApJ...244..205N} 
  Norris, J., Cottrell, P.~L., Freeman, K.~C., \& Da Costa, G.~S.\ 1981, 
  ApJ, 244, 205 
\bibitem[1999]{parmentier} Parmentier, G., Jehin, E., Magain, P.,
 Neuforge, C., Noels, A. \& Thoul, A. A., 1999, A\&A, 352, 138
\bibitem[1999]{powell}Powell, D. C., Iliadis, C., Champagne, A. E.,
  Grossmann, C. A., Hale, S. E., Hansper, V. Y., \& McLean, L. K., 
  1999, Nucl. Phys. A, 660, 349
\bibitem[Reimers(1975)]{1975MSRSL...8..369R} Reimers, D. 1975, Memoires of 
  the Societe Royale des Sciences de Liege, 8, 369
\bibitem[1999]{smith} Smith, G., 1999, PASP 762, 980
\bibitem[Smith \& Norris(1993)]{1993AJ....105..173S} Smith, G.~H.~\& 
  Norris, J.~E.\ 1993, AJ, 105, 173 
\bibitem[Suntzeff \& Smith(1991)]{1991ApJ...381..160S} Suntzeff, N.~B.~\& 
  Smith, V.~V.\ 1991, ApJ, 381, 160 
\bibitem[1991]{} Thielemann, F.,  Arnould, M. \& Truran, J.~W. 1991, 
  in Advances of Nuclear Astrophysics, ed. E. Vangioni-Flam et
  al. (France: Editions Fronti\'{e}res), 525
\bibitem[2002]{thoul} Thoul, et al. 2002, A\&A 383, 491
\bibitem[2002]{umeda} Umeda, H. \& Nomoto, K., 2002, ApJ 565, 385
\bibitem[Vassiliadis \& Wood(1993)]{1993ApJ...413..641V} Vassiliadis, E.~\& 
  Wood, P.~R.\ 1993, ApJ, 413, 641
\bibitem[Yong et al.(2003)]{2003A&A...402..985Y} Yong, D., Grundahl, F., 
  Lambert, D.~L., Nissen, P.~E., \& Shetrone, M.~D.\ 2003, A\&A, 402, 985 

\end{thebibliography}
\end{document}